\documentclass[a4paper,11pt]{article}
\pdfoutput=1
\usepackage{jheppub} 
\usepackage{graphicx,color,rotating}
\usepackage{hyperref}
\usepackage{epsfig,color}
\usepackage{slashed}
\usepackage{amsfonts}
\usepackage{tabularx}
\usepackage{graphicx}
\usepackage{adjustbox}
\usepackage{tabularx}

\begin{document}
\title{Gravitational wave signals of dark matter freeze-out}

\renewcommand{\thefootnote}{\arabic{footnote}}

\author{
Danny Marfatia$^{1}$ and
Po-Yan Tseng$^{2}$}
\affiliation{
$^1$ Department of Physics and Astronomy, University of Hawaii,
Honolulu, HI 96822, USA \\
$^2$ Department of Physics and IPAP, Yonsei University,
Seoul 03722, Republic of Korea \\
}
\date{\today}

\abstract{
We study the stochastic background of gravitational waves which accompany the sudden freeze-out of dark matter triggered by a cosmological first order phase transition that endows dark matter with mass. We consider models that produce the measured dark matter relic abundance via (1) bubble filtering, and (2) inflation and reheating, and show that gravitational waves from these
mechanisms are detectable at future interferometers.
}

\maketitle


\section{Introduction}

The identity of dark matter (DM) and its production mechanism 
are among the most important open questions in physics.
In the weakly interacting massive particle (WIMP) paradigm,
with its thermal freeze-out mechanism,
the measured DM relic density requires a WIMP mass of $\mathcal{O}(10\text{-}10^{3})$~GeV,
and an electroweak-scale DM annihilation cross section.
The decoupling temperature $T_{\rm dec}$ is related to the DM mass $m_\chi$ by
$T_{\rm dec}\simeq m_{\chi}/24$.
The vanilla version of this scenario had been challenged 
by the non-observation of WIMPs in DM direct detection searches.
Alternative scenarios for DM production in the early universe often assume
a DM sector that is out of thermal equilibrium with the standard model (SM) sector. For example, 
DM may be produced in the decays of a heavy particle~\cite{Kolb:1998ki,Allahverdi:2018iod}.
DM may also be produced by freeze-in through the feeble annihilation of particles which are thermalized with the SM 
bath~\cite{McDonald:2001vt,Hall:2009bx,Chu:2013jja}.
However, in all of the above scenarios, the DM mass is constant during DM production.

The discovery of the 125~GeV SM-like Higgs boson $h$ 
at the Large Hadron Collider (LHC)~\cite{Aad:2012tfa,Chatrchyan:2012xdj} consolidates 
spontaneous symmetry breaking as the mechanism that gives the SM particles their mass.
The Higgs mechanism gives the simple relation $m_f = y_f\cdot v_{\rm SM}$ 
between the fermion mass $m_f$
and its Yukawa coupling to the Higgs boson $y_f$, where $\langle h \rangle \equiv v_{\rm SM}\simeq 246$~GeV 
is the vacuum expectation value (VEV) of the SM Higgs. 
A picture of the universe going through an electroweak phase transition
because finite temperature effects modify its scalar potential 
as the universe cools down, emerges.
Before the phase transition, when all the SM particles are massless, the global minimum of the scalar potential is located at 
$\langle h \rangle=0$.
After the phase transition, the global minima of the potential shift to non-trivial values $\langle h \rangle \neq 0$, 
which gives mass to the SM particles.
In the SM, the electroweak phase transition is found non-perturbatively to be a smooth crossover~\cite{crossover1,crossover2}.
However, since we do not fully understand the entire structure 
of the scalar potential of the 125~GeV Higgs boson, and since the existence of additional scalars is a possibility,
 the nature of the transition is unknown.

The DM mass may be generated by a similar mechanism~\cite{Hambye:2013sna,Hambye:2018qjv,Heurtier:2019beu,Baker:2019ndr,Chway:2019kft}.
The mass originates from its couplings to a scalar, 
which obtains a non-trivial VEV in the early universe, so that massless DM becomes massive during the phase transition.
The scalar may or may not be the 125~GeV Higgs boson.
We consider a first order phase transition (FOPT)
in the early universe, with vacuum bubbles nucleated at temperature $T_\star$, 
which ends with the expanding bubbles populating the entire universe; until we discuss inflationary supercooling,
we do not differentiate between the nucleation temperature $T_n$
and the temperature $T_\star$ at which gravitational waves are produced.
The symmetric and broken phases are located outside and inside the bubbles, respectively.
The massless DM particles outside the bubbles become massive
when they enter the bubbles.
Only massless DM particles that
carry kinetic energy larger than $m_\chi$ 
can penetrate the bubble walls and become massive.
DM inside the bubbles abruptly decouples from the thermal bath if $T_\star < T_{\rm dec}$.
The result is that the bubbles filter out a
certain amount of DM and determine the 
DM relic abundance~\cite{Baker:2019ndr,Chway:2019kft}. The massless DM outside the bubbles remains thermalized with SM radiation. It is also possible that all the massless DM particles enter the bubbles after being diluted by a period of inflation, which determines the relic abundance~\cite{Hambye:2018qjv}.
DM particles with insufficient kinetic energy to enter the bubbles, bounce back to the symmetric phase and slow down the bubble expansion by applying pressure on the bubble walls.

The value of 
$m_\chi/T_\star$ needed to produce the correct DM relic abundance depends on
the velocity of the bubble walls $v_w$.
For instance, $T_\star\simeq m_\chi/30$ for $m_\chi=1$~TeV and 
$v_w = 0.01$, 
which satisfies $T_\star < T_{\rm dec}$.
Note that DM  freeze-out induced by a FOPT can easily accommodate DM masses above a  PeV,
which is beyond the current sensitivities of DM direct detection and LHC searches.

In this paper, we focus on gravitational wave (GW) signals 
of sudden DM freeze-out caused by a FOPT during which DM mass
is generated. 
Because the power and frequency spectrum of the GW signal is model dependent,
we choose two example models, $i)$ Scalar Quartic Model~\cite{Kehayias:2009tn,Wang:2020jrd,Dine:1992wr,Adams:1993zs} and 
$ii)$ $SU(2)_X$ Model~\cite{Hambye:2018qjv,Hambye:2013sna},
to demonstrate that in parameter space regions that  
yield the observed DM relic abundance, a detection is possible
at future GW interferometers.
In the Scalar Quartic Model, the
DM abundance is determined by bubble filtering, while in the $SU(2)_X$ Model, 
the DM abundance is set by inflation and reheating.

The paper is organized as follows. Bubble filtering is described in section~\ref{sec:bubble_filter},
and computations of the bubble wall velocity are detailed in section~\ref{sec:bubble_v}. 
In section~\ref{sec:GW}, 
we list the contributions to GW spectra from various processes. 
We calculate the GW signals for the two example models in section~\ref{sec:model},
and summarize in section~\ref{sec:summary}.

\bigskip

\section{Bubble filtering}
\label{sec:bubble_filter}


During the FOPT and bubble expansion, 
massless (massive) DM particles are located outside (inside) the bubble,
and momentum conservation must  be satisfied at the bubble wall.
An incident DM particle enters the bubble if it carries kinetic energy larger than its mass inside the bubble.
Otherwise, the massless DM particle is reflected and stays outside the bubble.
If a thermal flux of $\chi$ is incident on the wall, 
the number density of DM particles that enter
the bubble is~\cite{Chway:2019kft}
\begin{eqnarray}
\label{eq:filter_analytic}
n^{\rm in}_{\chi}=n^{\rm in}_{\bar{\chi}}
\simeq\frac{g_{\rm DM}T^3_\star}{\gamma_w v_w}\left(\frac 
{\gamma_w (1-v_w)m_\chi /T_\star+1}{4 \pi^2 \gamma^3_\omega(1-v_w)^2} 
\right)e^{-\frac{\gamma_w (1-v_w)m_\chi}{T_\star}}\,.
\end{eqnarray}
where $\gamma_w$ is the Lorentz boost factor of the wall in the rest frame of the plasma, $g_{\rm DM}$ is the number of spin states of the DM particle, 
and the DM distribution has been approximated to be Boltzmann.
%
%
In the non-relativistic limit, $v_w \to 0$, filtering strongly suppresses 
the DM number density inside the bubble as $e^{-m_\chi/T_\star}$.
In the relativistic limit, $m_\chi/(\gamma_w T_\star)\to 0$, the number density 
$\sim e^{-m_\chi/(2 \gamma_w T_\star)}$, so there is very little filtering and
$n^{\rm in}_\chi$
approaches the equilibrium number density outside the bubble, $n^{\rm eq}_{\chi}|_{T=T_\star} = g_{\rm DM}T^3_\star/\pi^2$.

If $T_\star$ is lower than the thermal decoupling temperature $T_{\rm dec}$, the DM inside the bubble 
is already decoupled from the thermal bath and makes up the DM relic abundance, 
On the other hand, if $T_\star> T_{\rm dec}$, the DM filtered by the bubble wall remains in
thermal equilibrium and the relic abundance is determined by standard thermal freeze-out with $m_\chi/T_{\rm dec}\simeq 24$.{\footnote{Note that even with the FOPT, $T_{\rm dec}$ is obtained
by equating the Hubble expansion rate $H$ and 
the thermal averaged DM annihilation rate, 
$\Gamma=\langle\sigma v\rangle n^{\rm in, eq}$~\cite{Baker:2019ndr}.
We assume that the SM  makes a dominant contribution to the light degrees of freedom  so that
$m_\chi/T_{\rm dec}\simeq 24$ 
with logarithmic corrections that depend on $m_\chi$, $T_\star$
and the DM coupling.}

The DM abundance today can be calculated by dividing  $n^{\rm in}_{\chi} + n^{\rm in}_{\bar{\chi}}$ (at $T_\star$) by the
entropy density $s=(2\pi^2/45)g_{\star S}T^3$, 
where $g_{\star S}$ is the effective number of relativistic degrees of freedom associated with entropy, and normalizing
to the critical density, $\rho_c = 3H^2_0 M^2_{\rm pl}$~\cite{Baker:2019ndr}:
\begin{eqnarray}
\label{eq:rescale}
\Omega_{\rm DM}h^2
%
%
\simeq 
6.29\times 10^{8}\, 
\frac{m_\chi (n^{\rm in}_\chi+n^{\rm in}_{\bar{\chi}})}{\rm GeV} \frac{1}{g_{\star S}T^3_\star}\,.
\end{eqnarray}
Using Eq.~(\ref{eq:filter_analytic}), this can be simplified to
\begin{eqnarray}
\label{eq:relic_apprx}
\Omega_{\rm DM}h^2\simeq \left\lbrace
\begin{array}{lc}
 1.27\times 10^8 
\left( \frac{m_\chi}{\rm GeV} \right)
\left( \frac{g_{\rm DM}}{g_{\star S}} \right)
\left( \frac{m_\chi}{2 \gamma_w T_\star}+1 \right) 
e^{-\frac{m_\chi}{2\gamma_w T_\star}}\,, &~~~ \text{for $v_w \to 1$}  \\
 3.19\times 10^7 
\left( \frac{m_\chi}{\rm GeV} \right)
\left( \frac{g_{\rm DM}}{g_{\star S}} \right)
\left( \frac{1}{v_w} \right)
\left( \frac{m_\chi}{T_\star}+1 \right) 
e^{-\frac{m_\chi}{T_\star}}\,, &~~~~ \text{for $v_w \to 0$.}
\end{array}
\right.
\end{eqnarray}
Then, $\Omega_{\rm DM}h^2\simeq 0.11$ requires
\begin{eqnarray}
\label{eq:relic_apprx1}
\begin{array}{lc}
\frac{m_\chi}{2\gamma_w T_\star} 
- \ln\left( \frac{m_\chi}{2\gamma_w T_\star} \right)
-\ln(g_{\rm DM})-\ln\left( \frac{m_\chi}{\rm GeV} \right) \simeq 16.2\,, &~~~\text{for $v_w \to 1$}  \\
\frac{m_\chi}{T_\star} -\ln\left(\frac{m_\chi}{T_\star}\right) +\ln(v_w) \simeq 22\,, &~~~~ \text{for $v_w \to 0$.}
\end{array}
\end{eqnarray}
For example, for $v_w\to 1$,  taking $m_\chi \approx1$ TeV and $g_{\rm DM}=2$, requires
\begin{equation}
\label{cond}
\frac{m_\chi}{2\gamma_w T_\star} \simeq 27\,,
 \end{equation}
  to give the measured DM relic abundance, 
$\Omega_{\rm DM}h^2\simeq 0.11$.

The DM relic abundances for three values of $T_\star$ and relativistic and non-relativistic wall velocities 
are shown in Fig.~\ref{fig:DM_relic}.
The left-panel shows that for small $v_w$,  $\Omega_{\rm DM}h^2 \simeq 0.11$ if $20 \lesssim m_\chi/T_\star \lesssim 40$ and
$1\,{\rm GeV}\lesssim T_\star \lesssim 1\,{\rm PeV}$. 
For $v_w \to 1$ (right panel), because bubble filtration is not efficient, larger values, 
$100 \lesssim m_\chi/T_\star \lesssim 170$, in
the exponent of Eq.~(\ref{eq:filter_analytic}) are needed
to suppress the DM number density. That larger $T_\star$ requires larger $m_\chi/T_\star$ 
can be understood by
combining Eq.~(\ref{eq:filter_analytic}) and~(\ref{eq:rescale}):
$
\Omega_{\rm DM}h^2 \propto T_\star \left(m_\chi / T_\star\right)^2 e^{-m_\chi/T_\star}
$.

\begin{figure}[t]
\centering
\includegraphics[height=2.9in,angle=270]{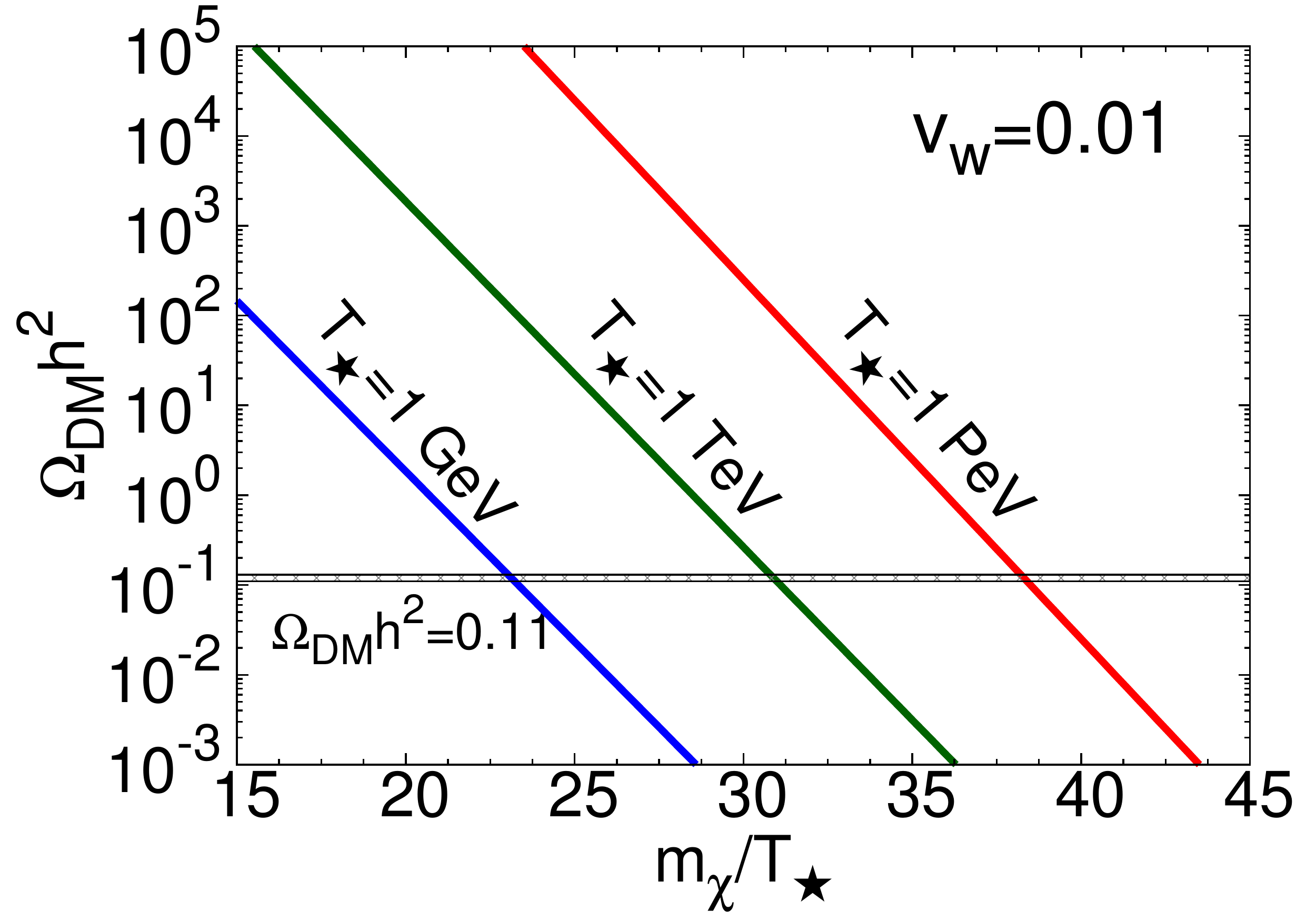}
\includegraphics[height=2.9in,angle=270]{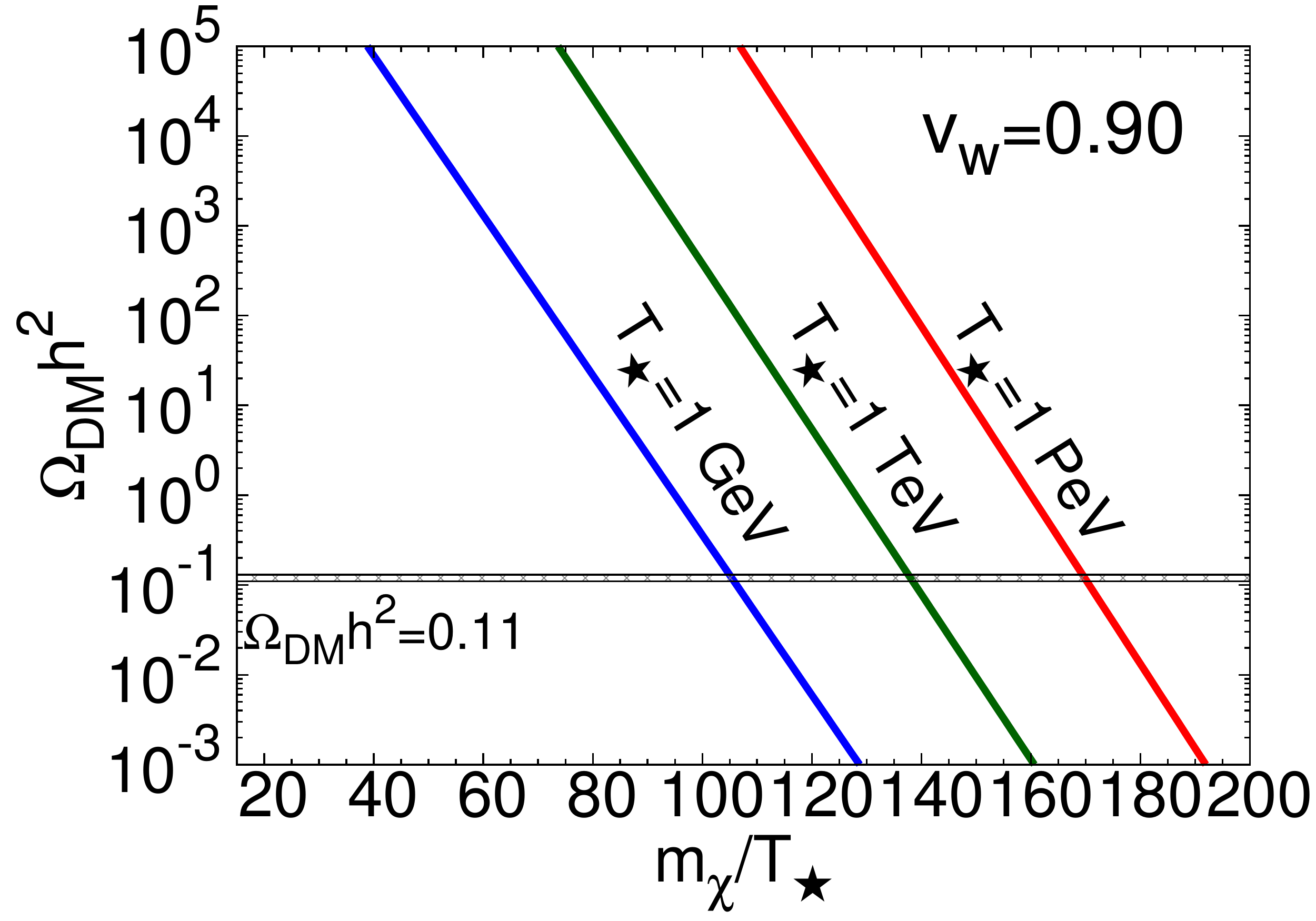}
\caption{\small \label{fig:DM_relic}
The DM relic abundance after bubble filtering for non-relativistic and relativistic  bubble wall velocities.
}
\end{figure}

\bigskip

\section{Bubble wall velocity}
\label{sec:bubble_v}

We consider a fermionic or bosonic DM particle $\chi$ that 
couples to a scalar $\eta$ (that could be the SM Higgs or a new particle) with coupling $g_\chi$ (not to be confused with $g_{\text{DM}}$, the number of spin states).
The scalar undergoes a
FOPT at temperature $T_\star$,
during which the VEV jumps from 
$\langle \eta \rangle=0$ to
$\langle \eta \rangle=v_\eta$.
%
Nucleation starts at $T_\star$,
and the bubbles expand and merge until the entire universe is
populated with the
$v_\eta$ phase.
During the bubble expansion two phases coexist. 
Inside the bubbles $\langle \eta \rangle = v_\eta$, 
and DM gets a mass $m_\chi \simeq g_\chi v_\eta$. Outside the bubbles, $\chi$ is massless because 
$\langle \eta \rangle = 0$. 
Bubble filtering occurs as described in the previous section.

DM particles that are reflected by the bubble wall
exert pressure $P$ on it, and slow down the bubble wall velocity, 
which is given by the equilibrium condition 
$\Delta V = P$,
where $\Delta V$ is the potential energy difference 
between the false and true vacua.
The strength of the phase transition is defined in terms of the latent heat of the transition,
\begin{eqnarray}
\label{eq:alpha}
\alpha \equiv \frac{\left(1-T\frac{\partial}{\partial T}\right)\Delta V|_{T=T_\star}}
{\rho_{\rm rad}(T_\star)}\,,
\end{eqnarray}
where the
radiation energy density, $ \rho_{\rm rad}(T)=\pi^2 g_\star T^4/30$, with $g_\star$ the number of effectively massless degrees of freedom at temperature $T$.
For the SM, far above the electroweak scale, $g_\star \simeq 106.75$.
Note that the derivative term in Eq.~(\ref{eq:alpha}) is negligible for
strong, supercooled transitions, as is the case 
for the $SU(2)_X$ Model.

In the ultrarelativistic limit, 
the pressure on the bubble wall can be obtained from the difference in the number of light degrees of freedom inside and outside the bubble~\cite{Chway:2019kft,Espinosa:2010hh,Bodeker:2009qy}:
$$
P=\frac{d_n g_\star\pi^2}{90}(1+v_w)^3 \gamma^2_\omega T_\star^4\,,
$$ 
where the  ratio of the number of light degrees of freedom is
$$
d_n \equiv \frac{1}{g_\star}\left[ \sum_{0.2 M_i > \gamma_w T_\star}\left(g^b_i+\frac{7}{8}g^f_i \right) \right]\,,
$$
with $g^b_i$ and $g^f_i$, the number of degrees of freedom of the bosons and fermions, respectively.
If particle $i$ of the thermal plasma gains mass $M_i$ inside the bubble and $0.2 M_i \gtrsim \gamma_w T_\star$,
then most of the $i$ particles fail to penetrate the wall
and instead exert pressure on it~\cite{Chway:2019kft}.
If $i$ is fermionic DM with $g_{\rm DM}=2$,
then $d_n\simeq 0.032$ including particle and antiparticle contributions.
Therefore, once $\alpha$ is known  
from the scalar potential, 
 $v_w$ can be obtained by solving the equation, $\Delta V = P$:
\begin{eqnarray}
\label{eq:V_P}
\alpha  =  \frac{d_n}{3}(1+v_w)^3 
\gamma^2_\omega\,.
\end{eqnarray}
%
%
In the limit $v_w \to 1$, with $d_n=0.032$, 
we find $\alpha \simeq 0.085 \gamma^2_\omega$ 
from Eq.~(\ref{eq:V_P}). Eliminating $\gamma_w$ from Eq.~(\ref{cond}) yields
%
the condition, 
\begin{equation}
\label{eq:relic_approx3}
\frac{m_\chi}{\sqrt{\alpha}T_\star}\simeq \frac{g_\chi v_\eta}{\sqrt{\alpha}T_\star} \simeq 185\,,
\end{equation}
to produce the measured relic abundance for $m_\chi \approx 1$~TeV.
If we assume
$g_\chi \simeq \mathcal{O}(1)$, a
large $v_\eta/T_\star \gtrsim \mathcal{O}(10)$
and small $\alpha \lesssim \mathcal{O}(0.1)$ is required.

A precise computation of the bubble wall velocity outside the ultrarelativistic 
regime is beyond the scope of this paper.
For bubble wall velocities faster than the speed of sound in the plasma ($1/\sqrt{3}$), but not ultrarelativistic,
we use the approximation~\cite{Steinhardt:1981ct},
\begin{eqnarray}
v_w=\frac{\frac{1}{\sqrt{3}}+\sqrt{\alpha^2+\frac{2}{3}\alpha}}{1+\alpha}\,.
\label{chap}
\end{eqnarray}
For $v_w \to 1/\sqrt{3}$ (i.e., $\alpha \to 0$), the condition for $\Omega_{\rm DM}h^2=0.11$ is
\begin{equation}
\frac{m_\chi}{T_\star}=\frac{g_\chi v_\eta}{T_\star}\simeq 75\,,
\label{rel}
\end{equation}
according to Eqs.~(\ref{eq:filter_analytic}) and (\ref{eq:rescale}).

The wall velocity in Eq.~(\ref{chap}) is fixed by the Chapman-Jouguet condition for fluid expansion in chemical combustion. Since this condition is generally not fulfilled~\cite{Laine:1993ey,Espinosa:2010hh}, 
we consider a range of velocities around the value of Eq.~(\ref{chap}) to parameterize the uncertainty in the predicted GW signal.

\bigskip

\bigskip

\section{Gravitational wave production}
\label{sec:GW}

A FOPT generates GWs from three processes~\cite{Caprini:2015zlo}:
i) Bubble collisions.
ii) Sound waves in the plasma following bubble collisions and before the kinetic energy is dissipated by bubble expansion. 
iii) Magnetohydrodynamic (MHD) turbulence in the plasma after the bubble collisions.
%
%
The parameters that control the signal are $v_w$, $T_\star$, the phase transition strength $\alpha$, the inverse of the duration of the phase transition $\beta/H_\star$ in units of the Hubble parameter at $T_\star$, 
all of which are model and scalar potential specific.

Our calculations of the GW spectra follow the semi-analytic treatment in Refs.~\cite{Huber:2008hg,Espinosa:2010hh,Caprini:2015zlo}. Here, we simply point out some 
aspects of the three contributions without regurgitating the equations used.
Increasing the values of $T_\star$ and $\beta/H_\star$ 
increases the peak GW frequency,
but the latter also suppresses the power of the GW signal.
The power also decreases as $v_w$ is decreased.  
These properties are shared by all three GW contributions.

The GW contribution from bubble collisions can be calculated directly from the scalar field $\eta$ in the envelope approximation.
In this approximation, an important quantity is the fraction of latent heat transformed into scalar field gradient energy, $\kappa_\eta$.

GWs are produced by the sound waves created during percolation.
For values of $v_w$ not too close to the sound speed or speed of light, parametric fits to the numerically obtained GW spectrum can be found in Ref.~\cite{Caprini:2015zlo}. These fits include an efficiency parameter $\kappa_v$ for the fraction of latent heat transformed into bulk motion of the fluid, that
depends on the expansion mode of the bubble.
The peak frequency of the contribution from sound waves is inversely propositional to $v_w$.

The contribution from MHD turbulence arises when
percolation  transfers a  $\kappa_{\rm turb}$
fraction of the latent heat into turbulence in the plasma.
This parameter is related to $\kappa_v$ via $\kappa_{\rm turb}\simeq \epsilon \kappa_v$, where 
$\epsilon$ represents the fraction of bulk motion that is turbulent.
The value of $\epsilon$ is still under investigation, 
and we conservatively take $\epsilon=0.05$~\cite{Caprini:2015zlo}, which makes the contribution from MHD turbulence small.

%
%

Even in the case of significant supercooling, as shown in Ref.~\cite{Bodeker:2017cim}, bubble walls do not runaway because of friction
provided by transition radiation. In our study of the $SU(2)_X$~Model, we explicitly check that the vacuum contribution driving the expansion dominates the pressure difference due to transition radiation across the wall. Then, the walls carry most of the energy and the GW signal arises from bubble collisions.

\bigskip

\bigskip

\section{Models}
\label{sec:model}

We now investigate two example models 
to demonstrate that abrupt DM freeze-out produces
a detectable stochastic GW background.

We consider the Scalar Quartic Model
and $SU(2)_X$ Model.
Both models have a quartic term as the highest order term in their scalar potentials.
However, in the former model, the effective 
scalar potential is composed of only one scalar field $\eta$, and may be viewed as approximating a multi-field potential.
There may be thermal or non-thermal contributions to the cubic term from new particles that are not heavy enough to  be integrated out~\cite{Wang:2020jrd}.
In this model, the DM candidate is unspecified.
On the other hand, the $SU(2)_X$ Model has the SM gauge group with an extra $SU(2)_X$, 
and the scalar potential at the Planck scale is assumed to only permit quartic 
terms built from the SM scalar doublet $H$ 
and a scalar doublet $S$ under $SU(2)_X$. The absence of quadratic terms renders the model dimensionless at tree level.
The quadratic terms and electroweak scale are dynamically generated~\cite{Heikinheimo:2013fta,Heikinheimo:2013xua},
and the $SU(2)_X$ vector bosons are automatically stable and are the DM candidates~\cite{Hambye:2008bq,Cirelli:2005uq}. A generalization of this model
that includes mass terms at tree level has been studied in Ref.~\cite{Baldes:2018emh}.

\bigskip
\subsection{Scalar Quartic Model}

The effective scalar potential at finite temperature is~\cite{Kehayias:2009tn,Wang:2020jrd,Dine:1992wr,Adams:1993zs}
\begin{eqnarray}
V_{\rm eff}(\eta,T) = \frac{\mu^2 +DT^2}{2} \eta^2-\xi T\eta^3 
+ \frac{\lambda}{4} \eta^4\,,
\end{eqnarray}
where we have neglected non-thermal contributions to the cubic term.
%
Many particle physics models such as the inert singlet, inert doublet, 
and minimal supersymmetry models, can be parametrized by the above finite-temperature effective potential. 
%
We identify $\eta$ with the SM Higgs and set the zero-temperature VEV to the SM value $v_\eta=v_{\rm SM}=246~{\rm GeV}$. Since
$\xi$ is not the Higgs trilinear coupling, but
the $T$-independent coefficient of the $\eta^3$ term 
of the high-$T$ expansion of the Higgs effective potential, we take $\xi$ as a free parameter.
Then the critical temperature is~\cite{Wang:2020jrd}
\begin{eqnarray}
T_c = \frac{2}{\lambda D - 2\xi^2} 
\left[ \frac{\sqrt{\lambda D(\lambda D - 2\xi^2)T^2_o)}}{2}  \right]\,,
\end{eqnarray}
where 
\begin{eqnarray}
T^2_0 = -\frac{\mu}{D} = \frac{\lambda}{D} v^2_{\rm SM}\,,
\end{eqnarray}
is the temperature when the potential barrier vanishes.
The two minima are 
\begin{eqnarray}
 \langle \eta \rangle = 0\,,~~~~\frac{3\xi T}{2\lambda} 
\left[1+ \sqrt{1-\frac{4\lambda (\mu^2+D T^2)}{9\xi^2 T^2}} \right]\equiv v_\eta \,.
\end{eqnarray}
There are three independent parameters $\xi,D$, and $\lambda$ in the above effective potential. 
For simplicity, we fix $\lambda=0.1$ in following analysis.
%

\begin{figure}[t]
\centering
\includegraphics[height=2.9in,angle=270]{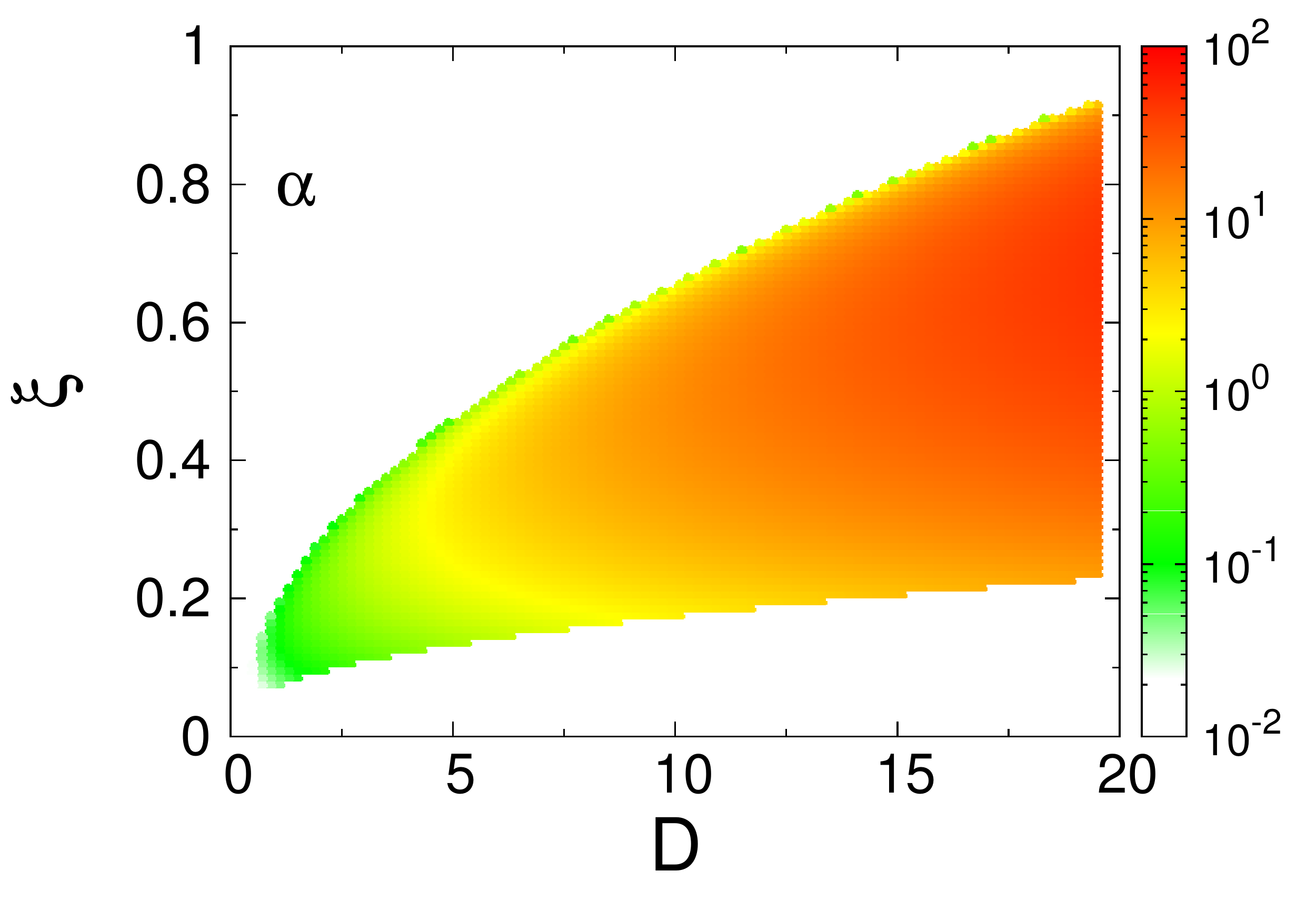}
\includegraphics[height=2.9in,angle=270]{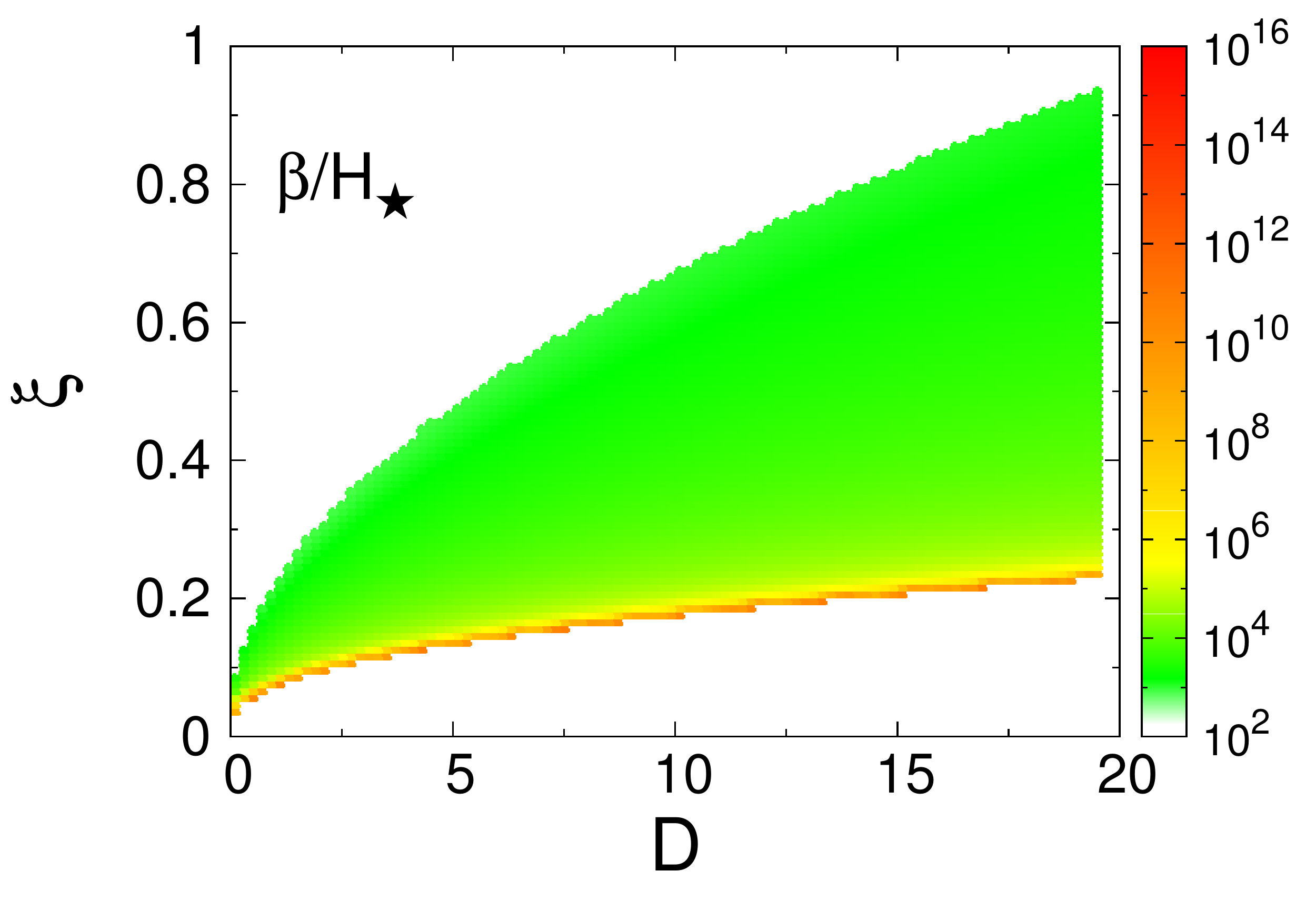}
\includegraphics[height=2.9in,angle=270]{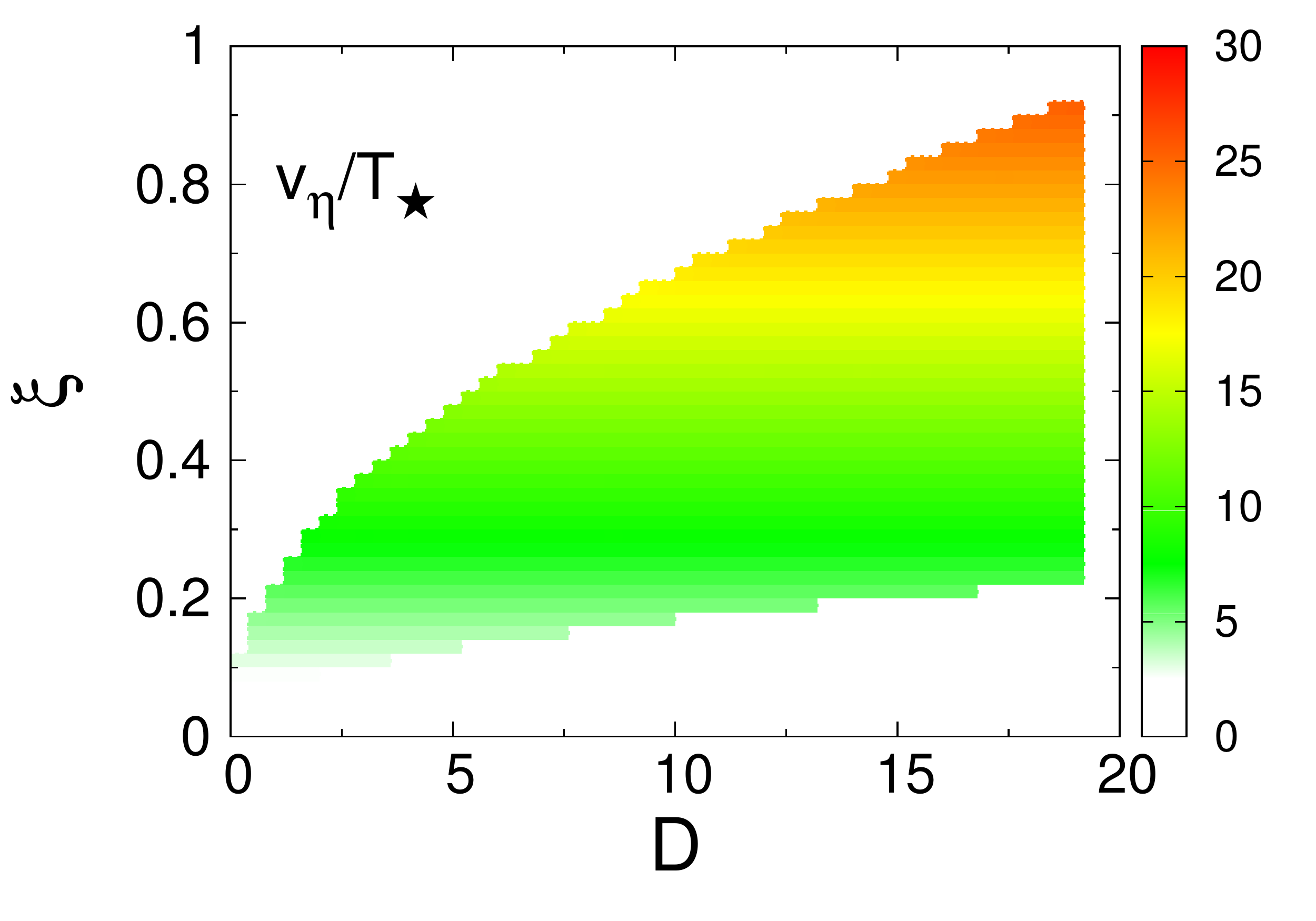}
\includegraphics[height=2.9in,angle=270]{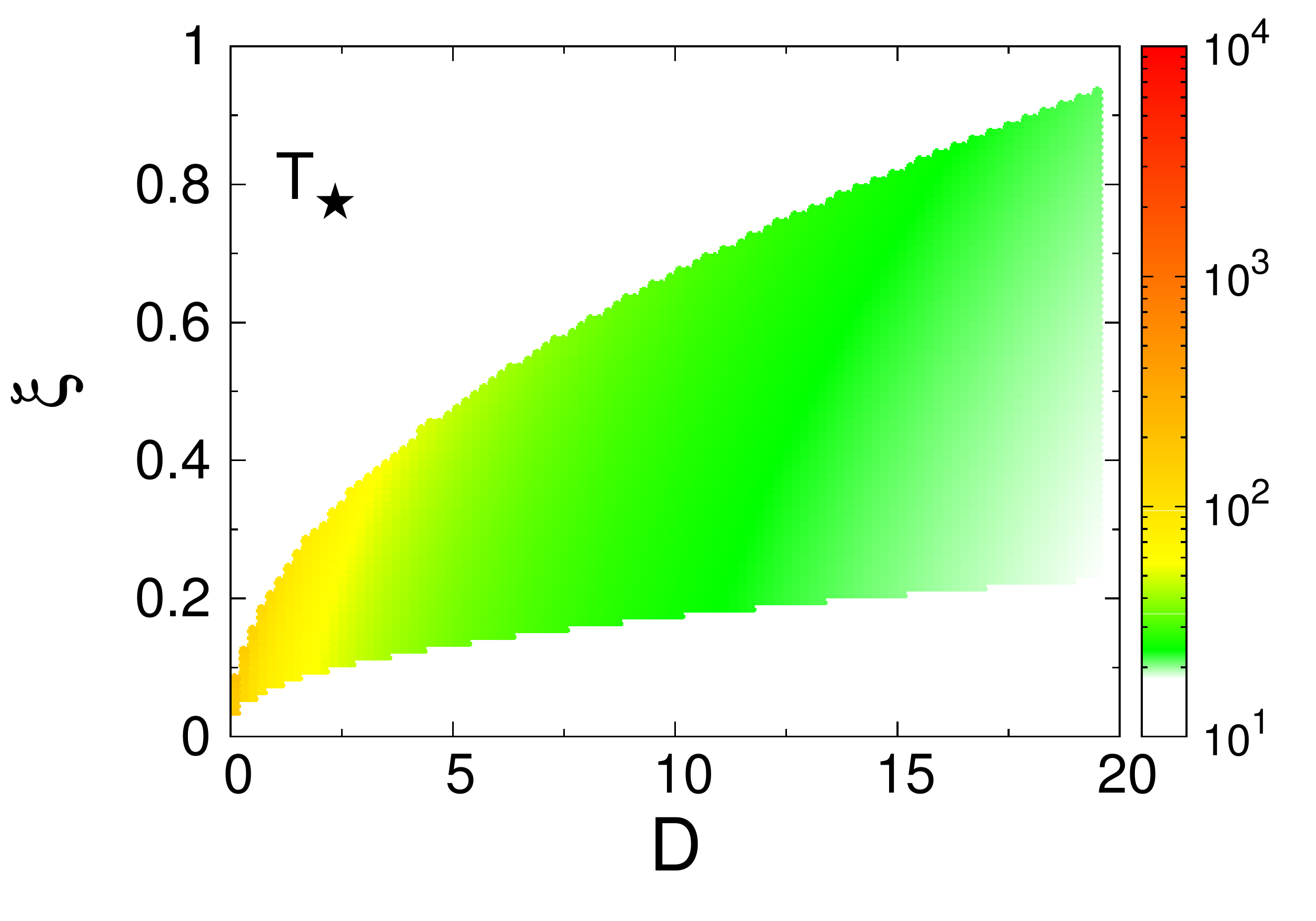}
\caption{\small \label{fig:quartic_GW_paramter}
The parameters $\alpha$, $\beta/H$, $v_\eta/T_\star$, 
and $T_\star$ for the Scalar Quartic Model  with $\lambda=0.1$.
}
\end{figure}

The nucleation temperature $T_n$ is determined by requiring 
the bounce action $S_3(T_n)/T_n \simeq 142$, 
when the vacuum tunneling rate equals the Hubble expansion rate~\cite{Kehayias:2009tn}.
We adopt the following analytic approximation from Ref.~\cite{Adams:1993zs}:
\begin{eqnarray}
\frac{S_3}{T}=\frac{64\sqrt{2}\pi}{81} \frac{\xi}{\lambda^{3/2}}
(2-\delta)^{-2}(\beta_1 \delta+\beta_2 \delta^2+\beta_3 \delta^3)\,,
\end{eqnarray}
where
$\delta\equiv \lambda (\mu^2+D T^2)/(\xi T)^2$, and
$\beta_1=8.2983$, $\beta_2=-5.5330$ and $\beta_3=0.8180$
are the  results of a numerical fit.
The expression is valid for $0<\delta <2$ which corresponds to
$T_0<T<T_c$.
We choose $T_\star = T_n$ to compute $\alpha$ 
and $\beta/H_\star = d(S_3/T)/d(\ln T)|_{T=T_n}$~\cite{Kehayias:2009tn}.
Figure~\ref{fig:quartic_GW_paramter} shows that in only narrow parameters region
(for example around $(D,\xi)\simeq (18,0.85)$), is $v_\eta/T_\star$ large enough ($\sim 25$)
as dictated by  Eq.~(\ref{rel}), to obtain the measured DM relic abundance for
$g_\chi\simeq \sqrt{4\pi}$.

\begin{figure}[t]
\centering
\includegraphics[height=2.9in,angle=270]{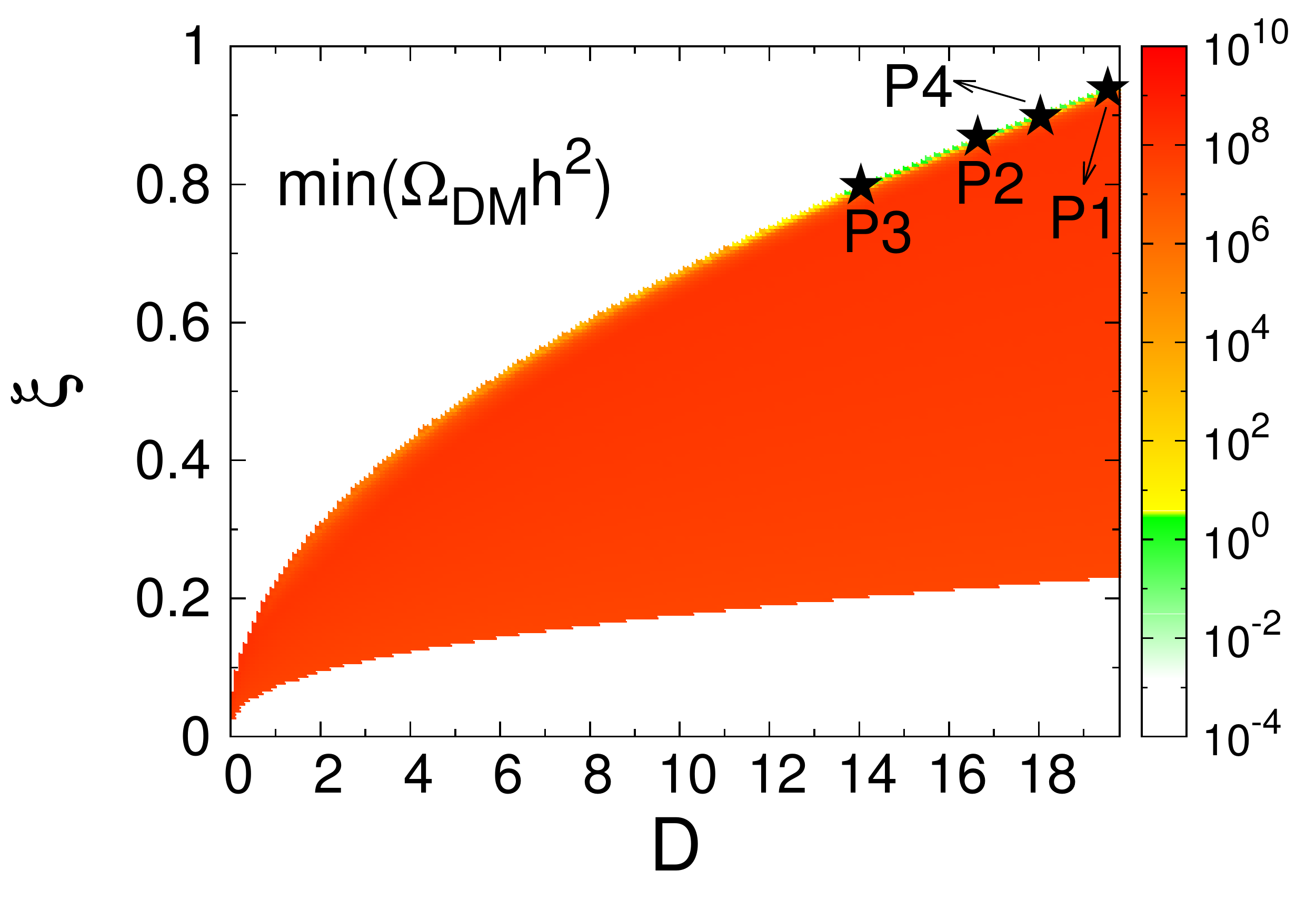}
\includegraphics[height=2.9in,angle=270]{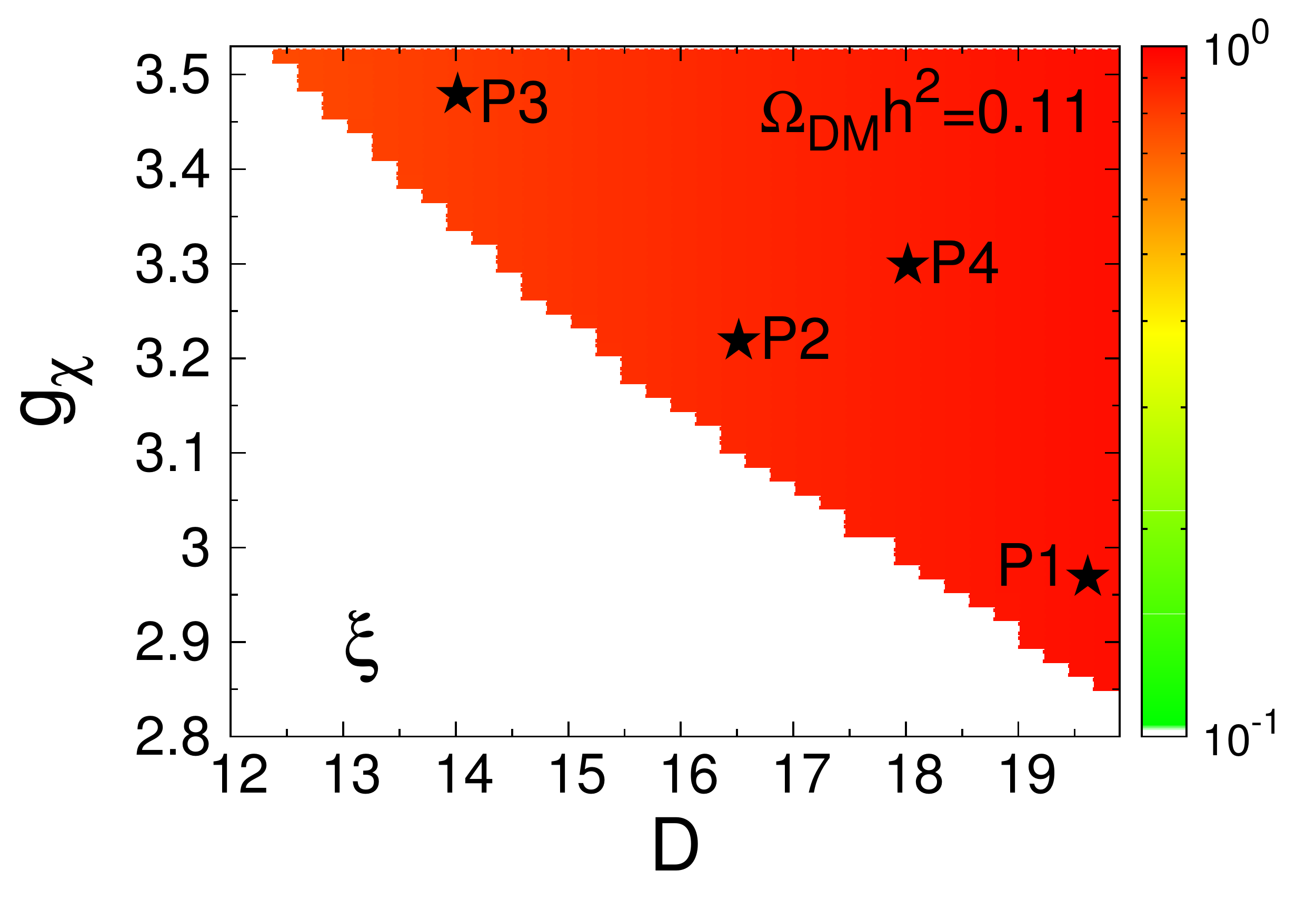}
\caption{\small \label{fig:quartic}
Scalar Quartic Model. Left: Minimal values of $\Omega_{\rm DM}h^2$ in the $(D,\xi)$ plane for $g_\chi \leq \sqrt{4\pi}$.
Right: Values of $\xi$ in the $(D,g_\chi)$ plane for $\Omega_{\rm DM}h^2=0.11$. 
The stars mark the four benchmark points in Table~\ref{tab:quartic_BP}. 
}
\end{figure}

The DM relic abundance is mainly determined by bubble filtering in the Scalar Quartic Model. 
Because of the presence of the quadratic term at tree-level, 
inflationary supercooling (as for the $SU(2)_X$ Model) does not occur.
In the left panel of Fig.~\ref{fig:quartic}, we show values of the relic abundance obtained  by varying $D$ and $\xi$
with $g_\chi \leq \sqrt{4 \pi}$.
In most of the parameter space DM is overproduced. 
However, in the narrow green region 
 $\alpha \lesssim 0.2$;
compare with the upper-left panel of Fig.~\ref{fig:quartic_GW_paramter}.
The values of $D,\xi$ and $g_\chi$ for which $\Omega_{\rm DM} h^2=0.11$ 
are displayed in the right panel of Fig.~\ref{fig:quartic}.
%
%
The four benchmark points marked with stars in Fig.~\ref{fig:quartic}
are listed in Table~\ref{tab:quartic_BP}.

The GW spectra for the benchmark points are shown in Fig.~\ref{fig:quartic_GW}.
The frequencies peak
around $\mathcal{O}(10^{-3}-10^{-2})$~Hz because $\beta/H_\star \simeq \mathcal{O}(1000)$
for all four points. This puts the model out of reach of LIGO and ET.
LISA, BBO and DECIGO are sensitive to all four benchmark points
because they have $\alpha\simeq 0.1$ which generates a large peak signal strength, $\Omega_{\rm GW}h^2 \sim 10^{-12}$. 

\begin{table}[t]
\caption{\small \label{tab:quartic_BP}
Benchmark points (with $\lambda=0.1$) for the Scalar Quartic Model  that give $\Omega_{\rm DM}h^2=0.11$.
}
\begin{adjustbox}{width=\textwidth}
\begin{tabular}{c|cccccc}
\hline
\hline
    & ~~~~~~{\bf P1}~~~~~~ & ~~~~~~{\bf P2}~~~~~~ & ~~~~~{\bf P3}~~~~~~ & ~~~~~~{\bf P4}~~~~~~  \\
\hline
$\xi$     
               & 0.943 & 0.863 & 0.796  & 0.901    \\
$D$     
               & 19.7 & 16.5 & 14.0 & 18.0    \\
$g_\chi$     
               & 2.97 & 3.22 & 3.48  & 3.31   \\
\hline
$\alpha$  & 0.089 & 0.082 & 0.076  & 0.121   \\
$\beta/H_\star$ & 1116 & 1062 & 1015  & 1085  \\
$v_\eta/T_\star$ 
          & 25.71 & 23.41 & 21.49 & 24.51  \\
$v_w$ & 0.768 & 0.763 & 0.760 & 0.791  \\
~~$T_\star/{\rm GeV}$~~   & 21.5 & 23.8 & 26.1 & 22.7   \\
~~$m_\chi/{\rm GeV}$~~ & 1642  & 1799  & 1953 & 1838    \\
\hline
\hline
\end{tabular}
\end{adjustbox}
\end{table}

\begin{figure}[t]
\centering
\includegraphics[height=3.8in,angle=0]{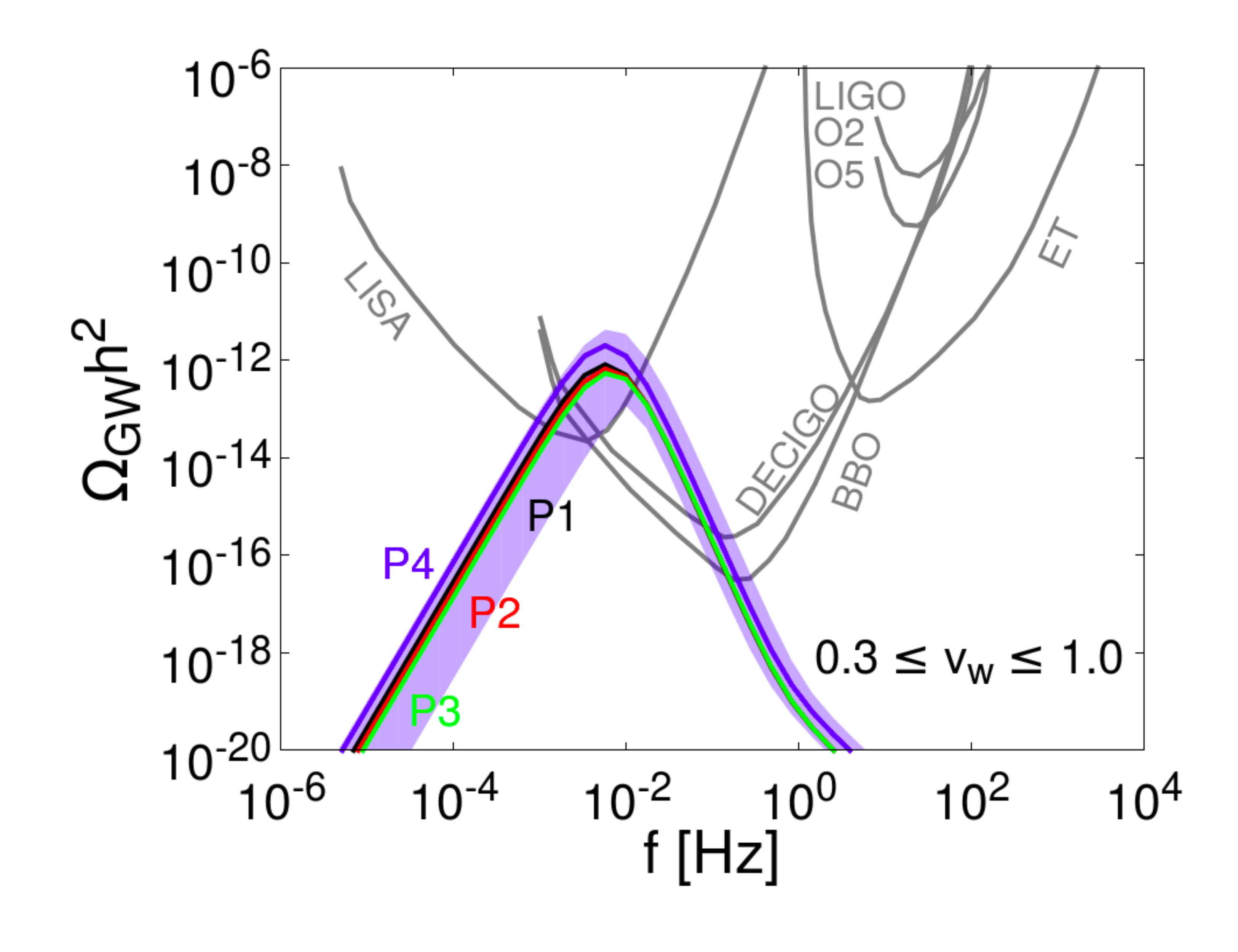}
\caption{\small \label{fig:quartic_GW}
The GW power spectra for the benchmark points of the Scalar Quartic Model in Table~\ref{tab:quartic_BP} and Fig.~\ref{fig:quartic}.
The shaded band straddling the {\bf P4} curve shows the uncertainty in the GW spectrum by varying the bubble wall velocity in the range, $0.3\leq v_w \leq 1.0$.
}
\end{figure}


\subsection{$SU(2)_X$ Model}

In this dimensionless model, the SM gauge group is extended by an $SU(2)_X$ with gauge coupling $g_X$,
and a scalar $S$, which transforms as a doublet under $SU(2)_X$ and is a singlet under the SM gauge group~\cite{Hambye:2018qjv,Hambye:2013sna}.
The scalar potential at tree level is
\begin{eqnarray}
\label{eq:lagrangian_su2x}
V &=& \lambda_H |H|^4-\lambda_{HS}|HS|^2+\lambda_S |S|^4, \nonumber \\
{\text {where~~~}} S &=& \frac{1}{\sqrt{2}}\left(
\begin{array}{c}
0 \\
\eta
\end{array}
\right),~~~~
H = \frac{1}{\sqrt{2}}\left(
\begin{array}{c}
0 \\
h
\end{array}
\right)\,.
\end{eqnarray} 
$SU(2)_X$ is spontaneously broken 
after $\eta$ acquires a VEV $\langle \eta \rangle= v_\eta$.
We treat the three vector bosons of $SU(2)_X$ cumulatively as a single DM candidate with $g_{\rm DM}=9$ 
and mass $m_\chi = g_X v_\eta/2$.

In this model, as the universe cools down, 
the universe remains trapped 
in the false vacuum 
(i.e., $\langle \eta \rangle=\langle h \rangle=0$) 
during thermal inflation due to the thermal effects.
Around this vacuum, all particles are massless.
When the energy of the false vacuum exceeds
the radiation energy (i.e., $\alpha > 1$), thermal inflation begins at temperature 
$T_{\rm infl}$ with Hubble constant $H_*$, which are given by
 \begin{eqnarray}
\frac{g_* \pi^2 T^4_{\rm infl}}{30}=\Delta V = \frac{3H_*^2 M^2_{\rm pl}}{8 \pi} \,.
\end{eqnarray}
During this phase, all particles 
undergo supercooling, because the 
scale factor grows exponentially
and the temperature falls inversely with the scale factor.
Supercooling ends at temperature $T_{\rm end}$
with a phase transition to the true vacuum at
$\langle \eta \rangle = v_\eta,\langle h \rangle = v_{\rm SM}$.
Supercooling ends when the temperature falls to the nucleation temperature $T_n$, or earlier at the QCD phase transition temperature $T_{\rm QCD}$ if $T_{\rm QCD}>T_n$:
\begin{eqnarray}
T_{\rm end}= \text{max}(T_n,T_{\rm QCD})\,,\quad
T_{\rm QCD}\simeq \frac{0.1\langle h \rangle_{\rm QCD}}{m_\chi/{\rm TeV}}\,,
\end{eqnarray}
where $\langle h \rangle_{\rm QCD} \simeq 100$~MeV.
%

The Coleman-Weinberg mechanism generates a true minimum at 
$\langle \eta \rangle = v_\eta$ when the quartic 
$\lambda_S$ becomes negative at a scale, $v_\eta e^{1/4}$~\cite{Hambye:2018qjv}. Assuming the true vacuum has zero energy, 
the energy in the false vacuum is $\Delta V\simeq 9m^4_\chi/(128 \pi^2)$, which implies
that supercooling starts at~\cite{Hambye:2018qjv}
\begin{eqnarray}
T_{\rm infl}\simeq \frac{m_\chi}{8.5}\quad\text{and}\quad
H_* = \sqrt{\frac{3}{\pi}}\,\frac{m^2_\chi}{4M_{\rm pl}}\,.
\end{eqnarray}
To compute the GW spectra we take
$T_\star=T_{\rm infl}$~\cite{Baldes:2018emh}.
To calculate $T_{\rm n}$, we use the bounce action,
\begin{eqnarray}
\label{eq:best_fit}
\frac{S_3}{T}= \left\lbrace
\begin{array}{l}
\frac{873.71}{g^{2.37}_X |\ln (0.60\, T/v_\eta)|}\,,~~~~~~~~~~~~~~~~~~~~~~~~\,\text{for $g_X< 1.18$} 
\\ [3mm]
142\times 
\frac{\ln(T_{\rm infl}/v_\eta)-e^{-4.7979(g_X-1.1779)}}{\ln(T/v_\eta)}\,,
~~~\text{for $g_X \geq 1.18$} 
\end{array}
\right.
\end{eqnarray}
which exactly reproduces the numerical result in Fig.~1 of Ref.~\cite{Hambye:2018qjv}.
Nucleation occurs when $S_3(T_n)/T_n\simeq 4 \ln (M_{\rm pl}/m_\chi)\simeq 142$.

After inflation ends,  the universe is reheated by the transfer of vacuum energy $\Delta V$ from the scalars to the other particles. How quickly this occurs determines 
the reheating temperature $T_{\rm rh}$. If the scalars decay rapidly, $T_{\rm rh} \sim T_{\rm infl}$, and if they oscillate and transfer energy at a rate $\Gamma$ much slower than the Hubble rate before decaying, $T_{\rm rh}$ is lower, i.e.,
\begin{equation}
T_{\rm rh} = T_{\rm infl} \ \text{min}\left(1, \Gamma/H \right)^{1/2}\,.
\end{equation}
We assume that the energy transfer rate is dominated by Higgs decay, 
so $\Gamma \simeq \Gamma_h \sin^2(v_{\rm SM}/v_\eta)$, where the Higgs decay width, $\Gamma_h \approx 4$~MeV.

\bigskip
\subsubsection{Dark matter abundance}

Having calculated $T_{\rm infl}$, $T_{\rm end}$
and $T_{\rm rh}$, we now consider the DM relic abundance in two regimes: 
$T_{\rm rh}> T_{\rm dec}$ and $T_{\rm rh}< T_{\rm dec}$,
where $T_{\rm dec}$ is the decoupling temperature in the conventional freeze-out scenario.
For $T_{\rm rh} < T_{\rm dec}$, the DM abundance is dictated by  
supercooling and by sub-thermal production via scattering.
Although we account for bubble filtering, its effect is negligible.
On the other hand, for $T_{\rm rh}> T_{\rm dec}$, the
supercooled population is washed out, and the sub-thermal population reattains thermal
equilibrium and produces the relic abundance as in the standard freeze-out scenario. 
The $\Omega_{\rm DM}h^2=0.11$ contour in the upper-left corner of Fig.~\ref{fig:su2x_Oh2} corresponds to this case.

The DM abundance resulting from inflationary supercooling is 
\begin{eqnarray}
\frac{n_{\rm DM}|_{T=T_{\rm rh}}}{s|_{T=T_{\rm rh}}}= \frac{45 g_{\rm DM}}{2\pi^4 g_*}\frac{T_{\rm rh}}{T_{\rm infl}} \left( \frac{T_{\rm end}}{T_{\rm infl}} \right)^3
\times f_{\rm in}\,,
\end{eqnarray}
where $f_{\rm in} \equiv (n^{\rm in}_\chi|_{T_{\rm end}})
/(n^{\rm eq}_\chi|_{T_{\rm end}})$ quantifies the filtering effect with $T_\star = T_{\rm end}$ in Eq.~(\ref{eq:filter_analytic}).
However, $f_{\rm in} =1$ for most of the parameter space of this model
because the bubble wall velocity 
is close to the speed of light
and $\gamma_w T_{\rm end} \gg m_{\chi}$.
The dilution from supercooling is significant 
for $T_{\rm infl}/T_{\rm end} \gg 1$, and can lead to DM being under-produced;
this corresponds to the white region in the lower-right corner 
of Fig.~\ref{fig:su2x_Oh2}.
The DM density today can be calculated by rescaling from $T_{\rm rh}$ to the temperature today, 0.235~meV, and using Eq.~(\ref{eq:rescale}).

\begin{figure}[t]
\centering
\includegraphics[height=3.8in,angle=270]{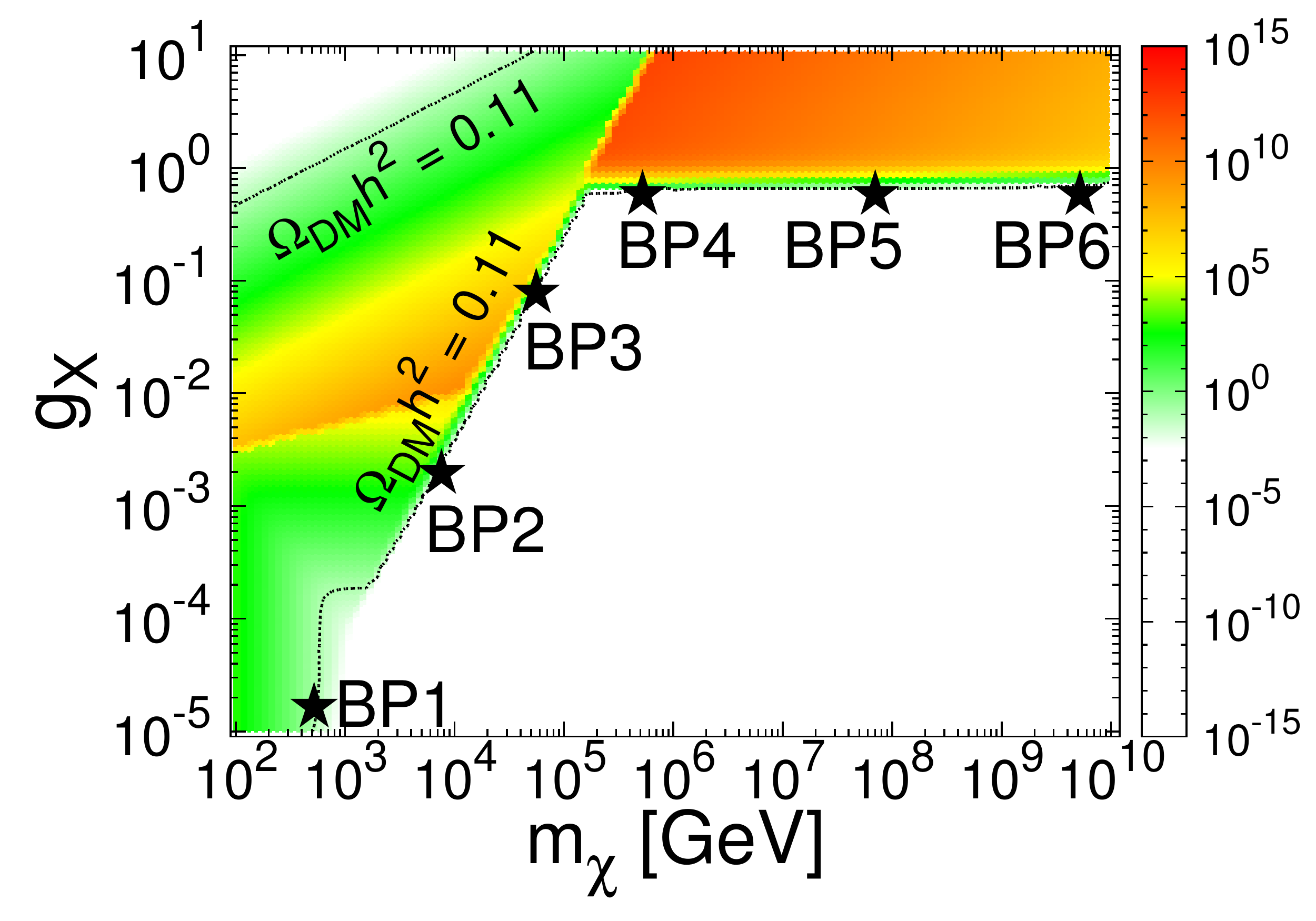}
\caption{\small \label{fig:su2x_Oh2}
 $\Omega_{\rm DM}h^2$ for the $SU(2)_X$ Model with $\langle h \rangle_{\rm QCD}=100~{\rm MeV}$.
The black-dashed contours indicate the observed DM relic abundance, $\Omega_{\rm DM}h^2=0.11$.
The stars mark the six benchmark points in Table~\ref{tab:su2x_BP}.
}
\end{figure}

We now consider sub-thermal DM production after supercooling. The decoupling temperature
of this population is $T_{\rm dec}\simeq m_\chi/\ln \lambda$, where 
$\lambda\equiv M_{\rm pl}m_\chi \langle \sigma_{\rm ann}v \rangle \sqrt{\pi g_\star/45}$, and
$\langle \sigma_{\rm ann}v \rangle$ is the thermal averaged DM annihilation cross section 
of the ${\rm DM}{\rm DM} \to \eta \eta$ 
process~\cite{Hambye:2018qjv}.
The abundance of the sub-thermal population is obtained by solving the Boltzmann equation.

For $T_{\rm rh} < T_{\rm dec}$, both supercooling and sub-thermal production contribute to the DM relic abundance,
\begin{eqnarray}
\Omega_{\rm DM}h^2=\Omega_{\rm DM}h^2|_{\rm supercool}
+\Omega_{\rm DM}h^2|_{\rm sub-thermal}\,.
\end{eqnarray}
For $T_{\rm rh} > T_{\rm dec}$, the plasma thermalizes again,  and the usual freeze-out mechanism yields the relic abundance,
\begin{eqnarray}
\Omega_{\rm DM}h^2=\Omega_{\rm DM}h^2|_{\rm freeze-out}\simeq
0.11 \times \frac{\langle \sigma_{\rm ann}v \rangle}{2\times 10^{-26}~{\rm cm^3/s}}\,.
\end{eqnarray}
The DM relic abundance
is shown in Fig.~\ref{fig:su2x_Oh2}.
We mark six benchmark points along the dashed curves (which indicate $\Omega_{\rm DM}h^2\simeq 
0.11$), 
and their values are listed in Table~\ref{tab:su2x_BP}.
For {\bf BP2} and {\bf BP3} sub-thermal processes dominate.
Dilution by supercooling fixes the DM abundances for 
{\bf BP1}, {\bf BP4}, {\bf BP5}, and {\bf BP6}.
The end of supercooling occurs at the nucleation temperature for  {\bf BP4}, {\bf BP5} and {\bf BP6}, and at the QCD phase transition temperature for {\bf BP1}.
 For the $\Omega_{\rm DM}h^2=0.11$ contour in the upper-left corner of Fig.~\ref{fig:su2x_Oh2}, $T_{\rm rh} > T_{\rm dec}$
 and the DM abundance is produced by the usual thermal freeze-out.
However, we have checked that the pressure difference due to transition radiation across the wall  becomes larger than the 
vacuum pressure driving the bubble expansion~\cite{Bodeker:2017cim}. Since this renders our assumption that
GWs are predominately produced by bubble collisions invalid, we do not consider this part of the parameter space any further.

\begin{table}[t]
\caption{\small \label{tab:su2x_BP}
Benchmark points for the $SU(2)_X$ Model that give $\Omega_{\rm DM}h^2=0.11$. 
Note that $T_\star=T_{\rm infl}$.
}
\begin{adjustbox}{width=\textwidth}
\begin{tabular}{c|cccccc}
\hline
\hline
    & {\bf BP1} & {\bf BP2} & {\bf BP3} & {\bf BP4} 
    & {\bf BP5} & {\bf BP6} \\
\hline
$m_\chi/{\rm GeV}$  & 540 & $5.4\times 10^3$ 
                    & $4.2\times 10^4$ & $3.0\times 10^5$ 
                    &  $8\times 10^7$   & $6\times 10^9$   \\
$g_X$               & $1.7\times 10^{-5}$ & $1.5\times 10^{-3}$ 
                    & $5.9\times 10^{-2}$ & $0.72$ 
                    &  $0.77$  & $0.82$  \\
\hline
$\alpha$  & $1.3\times 10^{14}$ & $1.2\times 10^{22}$ 
          & $1.8\times 10^{29}$ & $4.4\times 10^{16}$ 
          & $2.8\times 10^{13}$ & $9.4\times 10^{10}$  \\
$\beta/H_\star$ & $3.5\times 10^{11}$ & $8.1\times 10^6$ 
                & $1.3\times 10^3$ & 10.7 
                & 12.5 & 14.4  \\
$v_\eta/T_{\rm end}$ & $3.4\times 10^9$ & $3.7\times 10^9$ 
                     & $6.0\times 10^9$ & $3.4\times 10^{5}$ 
                     & $5.1\times 10^4$ & $1.2\times 10^4$  \\
$\gamma_w$ & $2.9\times 10^5$ & $2.9\times 10^7$ 
           & $1.8\times 10^9$ & $1.2\times 10^6$ 
           & $2.0\times 10^5$ & $4.7\times 10^4$  \\
$T_{\rm end}/{\rm GeV}$ & $1.85\times 10^{-2}$ & $1.87 \times 10^{-3}$ 
                    & $2.4\times 10^{-4}$ & 2.42 
                    & $4.01\times 10^3$ & $1.27\times 10^6$  \\
$T_{\rm rh}/{\rm GeV}$ & 46.6 & 422 
                       & 2082 & 3566 
                       & 14.5 & 0.201  \\
$T_{\rm infl}/{\rm GeV}$ & 63.6 & 629 
                    & $4.95\times 10^{3}$ & $3.53\times 10^{4}$ 
                    & $9.29\times 10^6$ & $7.08\times 10^8$  \\
\hline
\hline
\end{tabular}
\end{adjustbox}
\end{table}


\subsubsection{Gravitational wave signals}


To calculate the GW spectra, we need the phase transition strength $\alpha$, inverse phase transition duration $\beta/H_\star$, $v_w$, and $T_\star$.
We evaluate $\alpha$ and $v_w$ by following 
the procedure of Section~\ref{sec:bubble_v} and replacing
$T_\star$ by $T_{\rm end}$ in Eq.~(\ref{eq:alpha}) to make the equation valid for vacuum transitions~\cite{Caprini:2015zlo}.
The bounce action is used to find 
$\beta/H_\star = d(S_3/T)/d(\ln T)|_{T=T_{\rm end}}$.
We take $T_\star=T_{\rm infl}$, and rescale the peak frequency by $(T_{\rm rh}/T_{\rm infl})^{1/3}$ and the amplitude 
by $(T_{\rm rh}/T_{\rm infl})^{4/3}$ to account for a period of matter domination after the phase transition~\cite{Baldes:2018emh}.
The values of these parameters are provided in Table~\ref{tab:su2x_BP} for the six benchmark points.
The extremely large values of $\alpha$ and $\gamma_w$ are representative of ultra supercooling, 
for which the pressure $P$ cannot counter the vacuum energy $\Delta V$, 
so that bubble expansion keeps accelerating, 
until the bubbles collide and produce
GWs. All our benchmark points satisfy the requirement that $\Delta V$ is much larger than the pressure from transition radiation~\cite{Baldes:2018emh}:
\begin{equation}
T_{n}^4 \ll  {\beta \over H_\star} {m_\chi^5 \over g_X^2 M_{\rm pl}}\,.
\end{equation}

In Fig.~\ref{fig:su2x_GW}, we display the GW spectra for a few benchmark points and the sensitivities of 
the LIGO O2 and O5 observing runs~\cite{Aasi:2013wya}, 
LISA~\cite{Caprini:2015zlo,Auclair:2019wcv}, ET~\cite{Hild:2010id}, 
BBO~\cite{Yagi:2011wg}, 
and DECIGO~\cite{Kawamura:2019jqt} are} provided for comparison. 
The {\bf BP1}, {\bf BP2} and {\bf BP6} signals  are suppressed to an unobservable level
because of the 
large $\beta/H_\star$  for {\bf BP1} and {\bf BP2},  and small $T_{\rm rh}/T_{\rm infl}$ for {\bf BP6}.
{\bf BP3} and {\bf BP4} produce 
strong signals at BBO and DECIGO, and {\bf BP5} is marginally detectable at BBO.
{\bf BP4} can also be detected by LISA, and marginally by ET.


\begin{figure}[t]
\centering
\includegraphics[height=3.8in,angle=270]{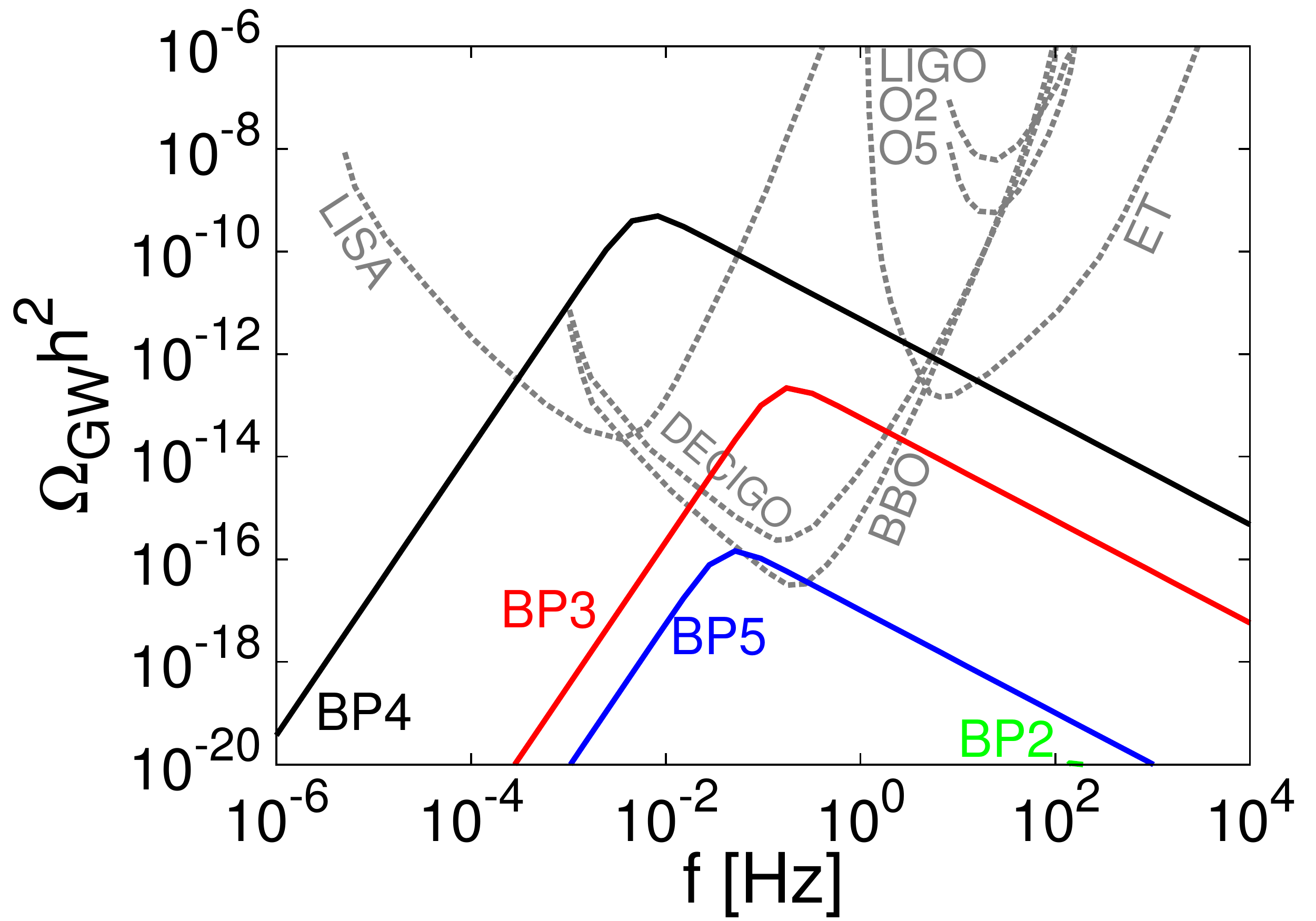}
\caption{\small \label{fig:su2x_GW}
The GW power spectra for benchmark points of the $SU(2)_X$ Model in Table~\ref{tab:su2x_BP} and Fig.~\ref{fig:su2x_Oh2}.
The signals of {\bf BP1} and {\bf BP6} are too small to display.
}
\end{figure}

\bigskip

\bigskip

\section{Summary}
\label{sec:summary}

We studied the sudden freeze-out of DM as an alternative to the continuous thermal freeze-out mechanism. 
A necessary ingredient for sudden freeze-out is that a FOPT generates DM mass. 
DM mass is generated via the coupling to a scalar particle, 
whose potential is responsible for a FOPT.
When the scalar field acquires a non-zero VEV, DM becomes massive. The DM relic abundance may be determined
by bubble filtering or by inflation and reheating.
Because a FOPT triggers sudden DM freeze-out, GWs offer a signature for sudden freeze-out not available for thermal freeze-out.   

To assess the viability of GWs as a signal of sudden freeze-out, we considered 
two example models that produce a DM abundance either by bubble-filtering (Scalar Quartic Model) or by inflation and reheating ($SU(2)_X$ Model).  We showed that
the observed DM relic abundance can be realized in these models with detectable  
GW signals  in future interferometers.

In the Scalar Quartic Model, the perturbativity condition, $g_\chi\lesssim \sqrt{4\pi}$, forces the preferred parameter space to have a
large $v_\eta/T_\star \gtrsim 20$ and small
phase transition strength, $\alpha \lesssim 0.2$.
To produce the DM relic abundance, 
the expanding bubbles must filter out most of 
the thermal DM in the symmetric phase 
via a large $m_\chi/T_\star$ 
and non-relativistic bubble wall velocity.
In these parameter regions the GW spectra have peak frequencies 
$\mathcal{O}(10^{-2})$~Hz, and powers large enough to be probed by LISA, DECIGO, and BBO.

In the $SU(2)_X$ Model, bubble filtering has a negligible effect on the DM number density, and the DM relic abundance
is governed either by supercooling during thermal inflation
or sub-thermal DM production.
The parameter regions that give the DM relic abundance favor $\alpha \gg 1$,
which corresponds to ultra supercooling.
Therefore, GWs originate
from bubble collisions.
Observable GW spectra have peak frequencies between about $10^{-3}$~Hz to 1~Hz, 
and enough power to be probed by
LISA, BBO, DECIGO and ET.
For {\bf BP3} and {\bf BP4}, 
the GW power is above $\Omega_{\rm GW}h^2 \simeq 10^{-13}$.


\section*{Acknowledgements}  
 We thank P.~Schwaller and an anonymous referee for useful comments. D.M. is supported in part by the U.S. DOE under Grant No. de-sc0010504.


\end{document}